\documentclass[%
 reprint,
 superscriptaddress,
 amsmath,amssymb,
 aps,
 prl,
]{revtex4-2}

\usepackage{graphicx}
\usepackage{dcolumn}
\usepackage{bm}
\usepackage{hyperref}
\usepackage{xcolor}
\usepackage[caption=false]{subfig}

\begin{document}

\title{Prediction of superconductivity in mass-asymmetric electron-hole bilayers}

\author{Luca M. Nashabeh}
\email{lmn2163@columbia.edu}
\affiliation{Department of Physics, Columbia University, New York, New York 10027, USA}
\author{Liang Fu}
\email{liangfu@mit.edu}
\affiliation{Department of Physics, Massachusetts Institute of Technology, Cambridge, Massachusetts 02139, USA}

\date{\today}

\begin{abstract}
We study density-balanced, mass-asymmetric electron–hole bilayers as a tunable platform for correlated quantum phases. With independent control of carrier density and interlayer separation, the system exhibits a rich phase diagram, including exciton condensates, Wigner crystals, and for large hole-to-electron mass ratios, an electron-liquid hole-crystal phase. This mixed phase is an analog of two-dimensional metallic hydrogen, featuring an electron liquid immersed in and coupled to a lattice of heavy holes. We show that acoustic plasmons mediate an attractive interaction between electrons, leading to BCS-type superconductivity at experimentally accessible parameters. The superconducting transition temperature is calculated from first principles, and experimental realization in van der Waals heterostructures is discussed.   
\end{abstract}

\maketitle

\textit{Introduction}---Electron-hole bilayer systems present a tunable platform 
for exploring various quantum phases of matter. Much attention has been focused on the physics of 
interlayer excitons, such as 
excitonic insulators \cite{jerome_excitonic_1967, qi_thermodynamic_2023, du_evidence_2017, qi_competition_2025, shao_quantum_2024, calman_indirect_2018} and exciton superfluids \cite{shevchenko_phase_1994, kellogg_vanishing_2004, eisenstein_new_1992, de_palo_excitonic_2002, eisenstein_boseeinstein_2004, das_gupta_experimental_2011, fowler-gerace_transport_2024, zhu_2d_2025}, with recent experiments having shown the existence of tightly bound excitons in transition metal dichalcogenide (TMD) based electron-hole bilayers \cite{sivan_coupled_1992, qi_perfect_2025, nguyen_perfect_2025, chui_quantum_2020, seamons_coulomb_2009, croxall_anomalous_2008, fedichkin_transport_2015, efimkin_anomalous_2016}. This material platform is notable for its tunability: the interlayer separation can be controlled by the thickness of an hBN spacer, while the charge densities on the top and bottom layers are separately tuned by top and bottom gates.   
Beyond exciton physics, this highly tunable system also promises a wealth of many-body phases, including the emergence of Wigner crystals and quantum states of trions        
\cite{dai_strong-coupling_2024, luo_pairing-based_2023, schleede_phase_2012, kaneko_exact-diagonalization_2013, tan_exciton_2005, wu_theory_2015, joglekar_wigner_2006,  sharma_ground_2016, karkkainen_electronhole_2004}.

In this Letter, we predict novel charge-ordered and superconducting states in density-balanced electron-hole bilayers with asymmetric electron and hole masses, which go beyond the known phases at equal masses. We show that at large mass asymmetry, the heavy particles (without loss of generality, we assume they are holes) form a Wigner crystal, while the light particles (electrons) form a Fermi liquid. This mixed phase, a quantum liquid of electrons immersed in a lattice of positive charges, bears a close analogy to solid hydrogen, with holes playing the role of the ions. Quantum fluctuation of the hole Wigner crystal gives rise to acoustic plasmons---the analog of phonons in solid hydrogen. Therefore, we expect that electron-plasmon coupling qualitatively similar to electron-phonon coupling gives rise to superconductivity similar to that of solid hydrogen \cite{ashcroft_metallic_1968, babaev_superconductor_2004, jaffe_superconductivity_1981, barbee_first-principles_1989, ashcroft_hydrogen_2004, mcmahon_properties_2012, richardson_high_1997, tenney_metallic_2021}. Importantly, while the nuclear mass is fixed at roughly $10^3$ times the electron mass, in 2D materials the effective mass ratio $m_h/m_e$ can be tuned over a wide range and can be made of order $10-100$, leading to very strong electron-plasmon coupling that boosts superconductivity.  

\begin{figure}[!t]
    \centering
    \includegraphics[width=0.9\linewidth]{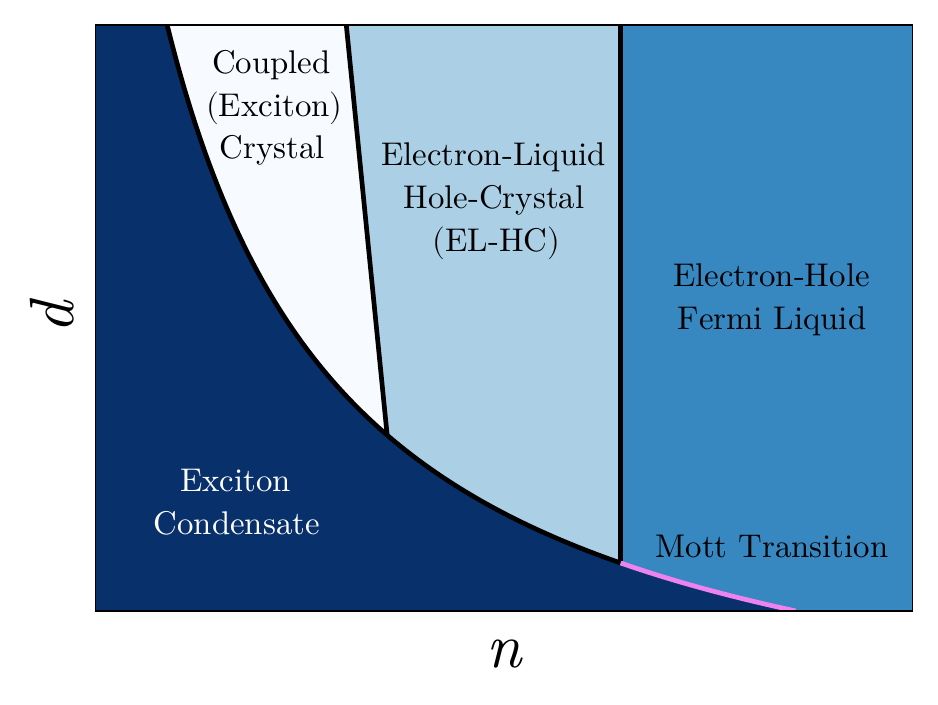}
    \caption{When $m_h/m_e>1$, the additional EL-HC phase emerges as compared to the symmetric bilayer. At sufficiently low interlayer distances, no crystallization occurs, and a Mott transition directly connects the exciton condensate and Fermi liquid phases. At larger interlayer distances, crystal phases become accessible, first in the hole and then in both layers.}
    \label{fig:schematic}
\end{figure}

\begin{figure*}[t]
    \centering
    \includegraphics[width=0.99\linewidth]{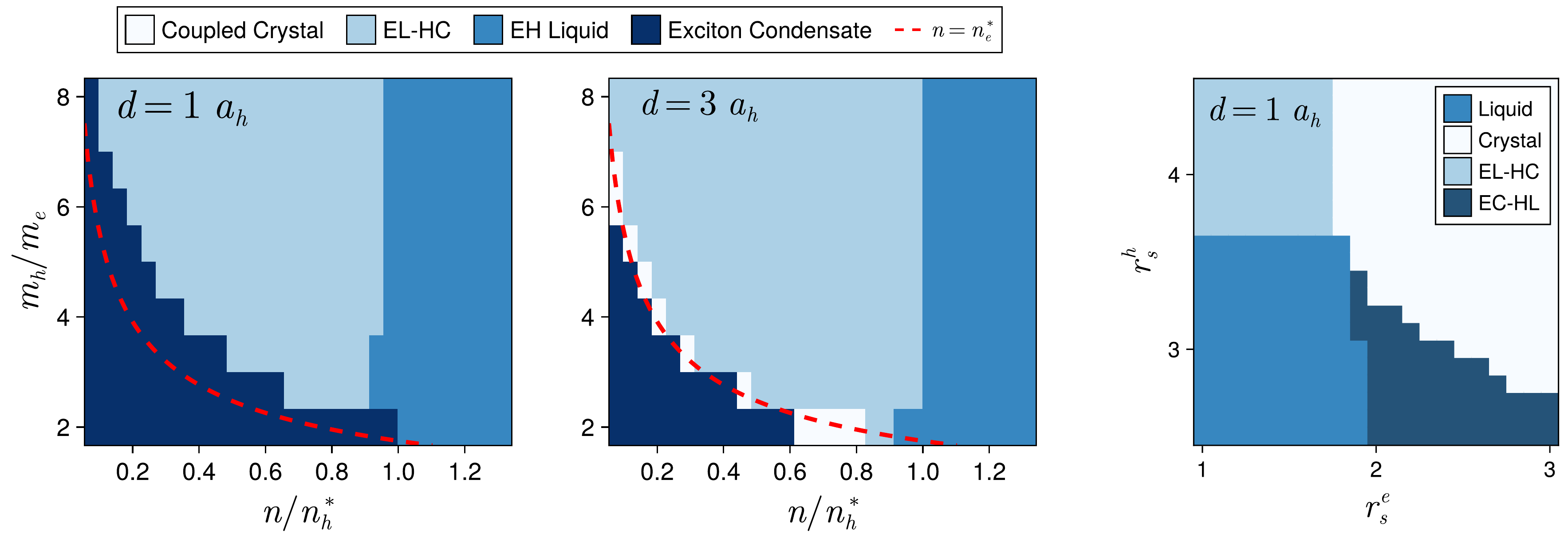}
    \caption{\textbf{Left:} The combined Hartree-Fock and Hartree-Fock-Bogoliubov phase diagram for $N_h=36$ spin-polarized holes (of fixed mass $m_h = m_0$ the bare mass) and equal electron populations of each spin. $n^*_h$ is the critical density for hole crystallization in an isolated monolayer. The dashed red line indicates the mass-dependent electron monolayer crystallization density $n=n_e^*$. Note the Mott transition between the Fermi liquid and exciton condensate phases for $d=1\, a_h$.
    \textbf{Right:} The pure Hartree-Fock phase diagram for the same scenario. This parameter range also includes the electron-crystal hole-liquid (EC-HL) phase.}
    \label{fig:phases}
\end{figure*}

Asymmetric bilayers feature a rich phase diagram controlled by three dimensionless parameters: the ratio of interlayer spacing to interparticle distance $\beta = d \sqrt{\pi n}$, with $n$ the electron or hole density; and the dimensionless Wigner-Seitz radii $r^e_{s} = (a_e\sqrt{\pi n})^{-1}$, $r^h_{s} = (a_h\sqrt{\pi n})^{-1} $, where $a_e = \hbar^2 \epsilon_r/(m_e e^2)$ and $a_h=\hbar^2 \epsilon_r/(m_h e^2)$ are the electron and n hole Bohr radii, which differ for unequal masses. $r_s^e$ and $r_s^h$ control the competition between kinetic and Coulomb interactions within each layer, while $\beta$ sets the scale for interlayer interactions. 

At zero temperature, we expect at least four phases to appear as the carrier density is varied by the applied interlayer potential bias, always keeping the net charge density zero. At low density and small interlayer spacing, the system is a dilute gas of interlayer excitons, which carry an out-of-plane dipole moment $e d$. The weak dipole-dipole repulsion between excitons results in an exciton superfluid. On the other hand, at very large layer spacing $d$ and intermediate density, strong intralayer Coulomb repulsion can lead to the formation of Wigner crystals in the different layers, which are weakly coupled by the attractive interlayer Coulomb interaction. As the density increases, each layer undergoes a quantum melting transition from a Wigner crystal into a Fermi liquid. Crucially, when the electron mass is smaller than the hole mass, there exists an intermediate density range where the hole $r^h_s$ is above the critical value for quantum melting but the electron $r^e_s$ is not. Then, we expect a new mixed phase with an electron liquid on one layer and a hole crystal on the other layer (EL-HC). A schematic phase diagram for asymmetric electron-hole bilayers is shown in Fig.~\ref{fig:schematic}, based on these general considerations. 

\textit{Phase Diagram}---To more quantitatively determine the phase diagram, we search for charge-ordered phases using Hartree-Fock and the condensate phase using Hartree-Fock-Bogoliubov (see Supplemental Material for details \cite{supp}). We present calculations at various fixed $d/a_h$. In the limit $d/a_h \gg 1$, the two layers partially decouple and independently transition from a liquid to crystal state at their respective critical densities. Note the two crystals are always aligned due to the attractive interlayer forces \cite{schleede_phase_2012, fogler_high-temperature_2014}. 
For intermediate $d/a_h$ however, we expect electron-hole interactions lead to richer phases. 

Our results from multi-species Hartree-Fock are shown in Fig.~\ref{fig:phases}. At larger interlayer spacings and densities, we see the expected non-condensate phases. Using a homogeneous Hartree-Fock-Bogoliubov ansatz, we also confirm that the formation of an exciton condensate is favored over the Fermi liquid and Wigner crystal states for smaller densities, with the critical density increasing at small interlayer separation.

To understand these phase diagrams, it is instructive to first consider the case of small mass asymmetry. If the masses are equal, the system exhibits three phases: a Fermi liquid, coupled crystal, or exciton condensate. The Fermi liquid and exciton condensate phases dominate at high and low densities respectively, while the coupled crystal can exist at intermediate densities and sufficiently large interlayer separations; otherwise, a Mott transition directly connects the liquid and condensate phases, as is visible for $d=1\;a_h$. As the mass asymmetry is increased---i.e. by decreasing the electron mass---the electrons tend to delocalize compared to the holes. 
Starting from either a crystal or condensate phase, this pushes the system towards the EL-HC phase, in which electrons form a liquid while the heavy holes are crystallized. 

Some additional observations can be made focusing only on the non-condensate phases, as shown on the right of Fig.~\ref{fig:phases}. At $d/a_h=1$, we see that the electron and hole crystallization depend on both $r_{s}^e$ and $r_{s}^h$, with electron crystallization promoted at higher values of $r_{s}^h$, consistent with previous PIMC studies \cite{schleede_phase_2012}. This can be understood from the interlayer Hartree potential $V^e_H \propto -|\psi_h|^2$ acting on the electrons. Since the depth of this potential increases with $r_{s}^h$ even after the hole crystal is formed, the critical electron crystallization density will also be enhanced. In our Hartree-Fock calculations, we only see this effect for small interlayer distances $d\lesssim a_h$. For $d/a_h \gtrsim 3$, the layers are still coupled, but crystallization in either depends only on its own density. 

These results are generally consistent with prior numerical studies \cite{ludwig_crystallization_2007, schleede_phase_2012, luo_pairing-based_2023}. 
Also, since it is the reduced mass $m_e m_h/(m_e+m_h)$ that appears in the condensate free energy, the condensation fraction is highest for symmetric masses, agreeing with other works \cite{kaneko_exact-diagonalization_2013, tan_exciton_2005, zhu_2d_2025}.  
However, previous studies neglect electron/hole spin, Fermi statistics, or both, and are thus inadequate for studying the intermediate phases of interest.  

\begin{figure}[!t]
    \centering
    \includegraphics[width=0.95\linewidth]{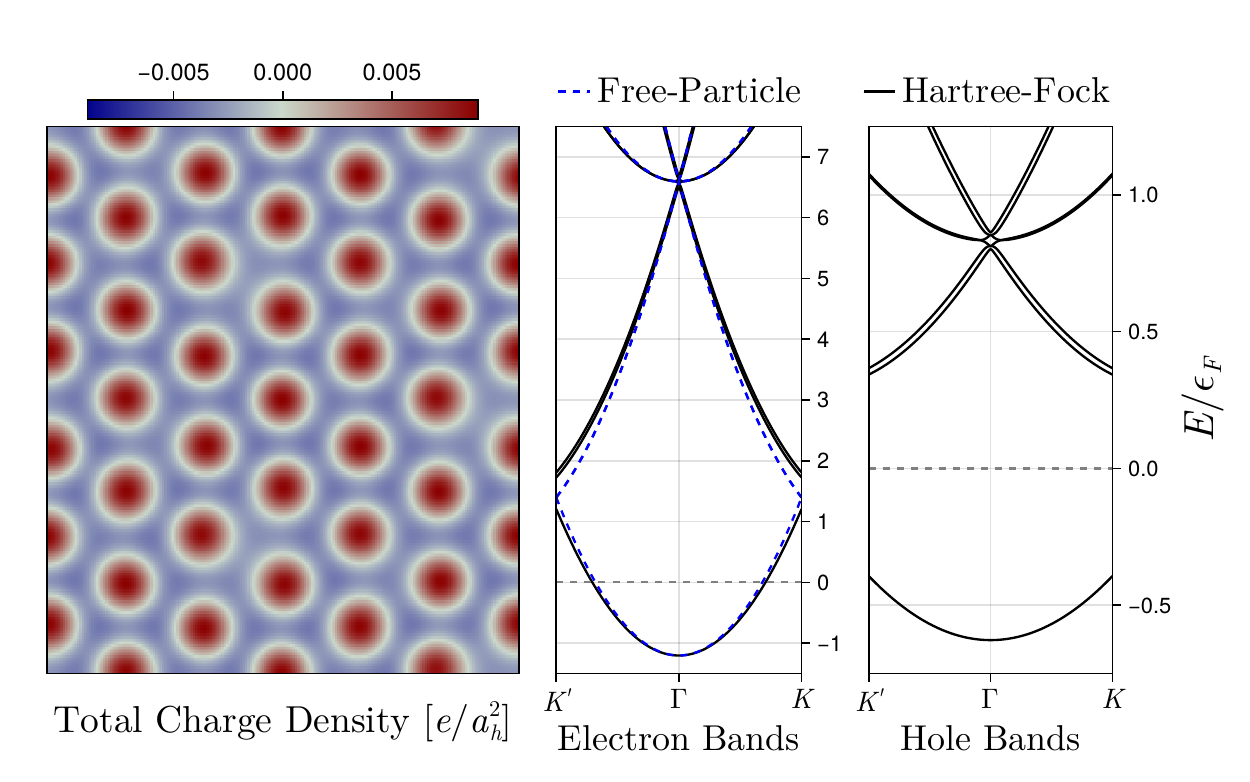}
    \caption{The restricted Hartree-Fock EL-HC phase for $d=a_h$ and $m_h/m_e = 10$ with 36 holes ($m_h = m_0$) at $r_s^h = 7$. $\epsilon_F$ is the free electron Fermi energy. The charge density shows the crystallized holes and liquid electrons. The Hartree-Fock electron band structure---referenced to the Fermi level---is well described by a free-particle model with $m_e^* \approx 0.93\, m_e$. The hole band structure has a gap $\Delta \approx 0.74\, \epsilon_F$ at the Fermi level.}
    \label{fig:density}
\end{figure}

\textit{Electron-Liquid Hole-Crystal}---Our Hartree-Fock calculations show the EL-HC phase to appear robustly for mass asymmetry $m_h/m_e \gtrsim 2$. 
A typical charge density pattern from this phase is given in Fig.~\ref{fig:density}. As expected, the holes assemble into a triangular Wigner crystal, while the electrons form a nearly homogeneous background liquid. We assume for simplicity that the crystallized holes are spin polarized (noting that the exchange energy between localized spins is very small), while retaining the spin-degeneracy of electrons forming the Fermi liquid.

Additional insight can be gleaned from the Hartree-Fock band structures of the electrons and holes in this phase. 
The electron band structure is very well described by a free-particle Hamiltonian; its dispersive band structure arises from the weak scattering off the electrostatic potential $V^e_H$. Moreover, the occupied band is only half-filled because of the spin freedom. 
On the other hand, the hole band structure fundamentally differs from the non-interacting limit, developing a gap at the Fermi level. This reflects the insulating character of a Wigner crystal. 

While the Hartree-Fock calculation qualitatively captures the EL-HC phase, it misses important correlation effects from electron-hole coupling which may lead to superconductivity, as we next address.

\textit{Theory of Plasmon-Mediated Superconductivity}---
Standard BCS theory predicts that superconductivity arises from attractive phonon-mediated interactions between electrons \cite{bardeen_theory_1957}. 
Given the close analogy to metallic hydrogen---with the heavy holes in the bilayer playing the role of the heavy nuclei---we now consider the possibility of superconductivity in the asymmetric bilayer. 

In the limit $d= 0$ and $m_h/m_e \approx 2000$, the analogy of our system with 2D hydrogen is exact, with superconductivity mediated by standard acoustic phonons. 
In metals, the linear dispersion for these acoustic phonon $\omega_{\bf q}= v_s |{\bf q}|$ is due to the coupled oscillation of electrons and ions where electrons screen the ion motion. Crucially, a proper treatment requires treating long-range Coulomb interactions between all charged particles on equal footing \cite{tupitsyn_coulomb_2016}. This same consideration applies to our bilayer system composed of electrons and holes. 

This requires special attention when $d\neq 0$, as the electron and hole layers are spatially separated. In the extreme limit $d\rightarrow \infty$, density oscillations in the two layers become decoupled, which has no counterpart in conventional metals and precludes acoustic phonons. At finite $d$ however, coupled oscillations still exist, but are more aptly described as plasmons, since there is a local charge modulation in each layer. Plasmons in bilayer systems have been studied extensively \cite{santoro_acoustic_1988, park_acoustic_2010, diaconescu_low-energy_2007, vazifehshenas_temperature_2010, das_sarma_plasmons_1998, vazifehshenas_temperature_2010, lozovik_electron-hole_2008, shao_electromagnetic_2025}. 
Indeed, there exists an acoustic plasmon mode, a collective oscillation where electrons and holes in the two layers move together, maintaining net charge neutrality. This mode has linear dispersion at small $q$, acting like an acoustic phonon. 



Having identified acoustic plasmons as the analog of phonons in metallic hydrogen, we now follow the method of Pines \cite{pines_superconductivity_1958} to estimate the effect of plasmon-mediated interactions on the net electron-electron potential. As the electron layer is most relevant for superconductivity, from here on most quantities are referenced with respect to the electron Bohr radius $a_e$ and Wigner-Seitz radius $r_s^e$. The Fermi surface---and associated quantities---refers to the electron's Fermi surface.

We approximate the direct electron-electron interaction with a 2D Thomas-Fermi screened intralayer potential, and assume the electron-plasmon coupling is solely determined by a similar interaction with an interlayer suppression. We also assume the mass ratio $m_h/m_e \gg 1$ is large enough that the holes may be treated like heavy ions. This gives Coulomb and electron-plasmon couplings
\begin{align}\label{eq:V_base}
    U({\bf k} - {\bf k}') &= \frac{2\pi e^2}{\epsilon_r} \frac{1}{|{\bf k} - {\bf k}'| + k_s} \\
    M_{{\bf q}}({\bf k} - {\bf k}') &= i  \sqrt{\frac{n \hbar}{2m_h \omega_{{\bf q}}}}\; \frac{2\pi e^2}{\epsilon_r} \frac{|{\bf k} - {\bf k}'| \;e^{-d|{\bf k} - {\bf k}'|}}{|{\bf k} - {\bf k}'| + k_s}
\end{align}
where ${\bf q} \equiv {\bf k} - {\bf k}'$ modulo the reciprocal lattice and lies in the first Brillouin zone, $\omega_{{\bf q}} = v_p |{\bf q}|$ with $v_p$ the acoustic plasmon velocity, and $k_s = 2\pi e^2 g_0/\epsilon_r = 2/a_e$ ($g_0$ is the density of states) is the Thomas-Fermi screening wave vector for a 2D system with parabolic dispersion.  

Assuming only electrons near the Fermi surface pair, the average effective pairing interaction is 
\begin{equation}\label{eq:V_avg}
    \begin{split}
        V &= \frac{1}{2} \left\langle U({\bf k} - {\bf k}') - \frac{2}{\hbar \omega_{{\bf q}}}|M_{{\bf q}}({\bf k} - {\bf k}')|^2\right\rangle\\
        &= \frac{\pi e^2}{\epsilon_r} \left\langle\frac{1}{\tilde k + k_s}- \frac{k_s \tilde k^2}{\alpha^2 q^2}\, \frac{ \,e^{-2d\tilde k}}{(\tilde k + k_s)^2}\right\rangle
    \end{split}
\end{equation}
where $\tilde k = |{\bf k} - {\bf k'}| = 2k_F |\sin(\phi/2)|$ and we average over the Fermi surface. $\alpha = v_p/v_0$ is the ratio between the acoustic plasmon velocity $v_p$ and the 2D Bohm-Staver sound velocity $v_0^2 = \epsilon_F/m_h$ \cite{ashcroft_metallic_1968}. $v_0$ is an approximation assuming perfect screening---electrons and holes move to maintain net charge neutrality, with the restoring force from the electron's Fermi pressure and the inertia from the hole mass.

To continue, we divide the average into two regimes. In the normal regime, ${\bf q} = {\bf k}-{\bf k}'$ and we can directly integrate. In the Umklapp regime ${\bf q} = {\bf k}-{\bf k}' + {\bf G}$ for some reciprocal lattice vector; to give an underestimate of the Umklapp effects, we simply set $q = k_D $. In total

\begin{widetext}
    \begin{equation}
        g_0 V = \frac{1}{\pi}\left[\int_0^{\pi/2} \frac{a }{\sin(\theta) + a}\,\mathrm{d}\theta - \frac{a^2}{\alpha^2} \left(\int_0^{\pi/4} \frac{e^{-4\sqrt{2}\beta \sin(\theta)}}{(\sin(\theta) + a)^2}\,\mathrm{d}\theta + \int_{\pi/4}^{\pi/2} \frac{2e^{-4\sqrt{2}\beta \sin(\theta)}}{(1+ a\csc(\theta))^2}\,\mathrm{d}\theta\right)\right] 
        \label{eq:super}
    \end{equation}
\end{widetext}
where $a= k_s/(2k_F ) = r_s^e/\sqrt{2}$. 
Additional derivational details are included in the Supplemental Material \cite{supp}. 

The BCS critical temperature is then given by the standard expression $T_c \approx 1.13\, T_D e^{1/(g_0 V)}$ \cite{bardeen_theory_1957}, where we may estimate the Debye temperature as
\begin{equation}
    T_D = \frac{2 \alpha\epsilon_F}{k_B} \sqrt{\frac{m_e}{m_h}}\label{eq:debye}
\end{equation}
The absolute value of $T_c$ depends on all 4 dimensionful quantities in the system: the density $n$, both masses $m_e$ and $m_h$, and the interlayer distance $d$. However, if we express $T_c$ in natural units of the electron Hartree temperature $T_H^e = m_e e^4/(\epsilon_r^2 \hbar^2 k_B)$, the ratio $T_c/T_H^e$ only depends on the three dimensionless ratios defining the phase space, i.e. $m_h/m_e$, $r_s^e$, and $d/a_e$. In particular, for fixed $d/a_e$, the \textit{maximum} critical temperature can only be a function of the mass ratio $m_h/m_e$.

\textit{Superconductivity Predictions}---As is clear from Eq.~\ref{eq:super}, the value of $\alpha$---the acoustic plasmon velocity in units of 2D Bohm-Staver sound velocity---is critical for predicting superconductivity; for small $\alpha$, the attractive plasmon-mediated interaction is significantly larger. Often, including for the alkali metals, $\alpha \approx 1$, but metallic hydrogen is predicted to have $\alpha \approx 0.45$, allowing superconductivity \cite{ashcroft_metallic_1968, ashcroft_compressibility_1967}.

The acoustic plasmon velocity can be related to the bulk modulus $B = A\; \partial^2 \!E/\partial^2\! A$ ($E$ is ground state energy) and mass density $\rho$ by the general relation $v^2 = B/\rho$. Since the bulk modulus is related to the electronic compressibility by $\kappa = \partial n/\partial \mu = n^2/B$, 
$v$ can be determined from the compressibility 
$v=  n/\sqrt{ \rho \kappa}$. 
Then, $\alpha$ can be expressed as a ratio of compressibilities 
\begin{equation}
    \alpha = \sqrt{B/B_0} = \sqrt{\kappa_0/\kappa}
\end{equation}
with $\kappa_0$ and $\kappa$ the compressibilities of the non-interacting electron gas used in the Bohm-Staver approximation and of the interacting system, respectively. 

\begin{figure}[!t]
    \centering
    \includegraphics[width=\linewidth]{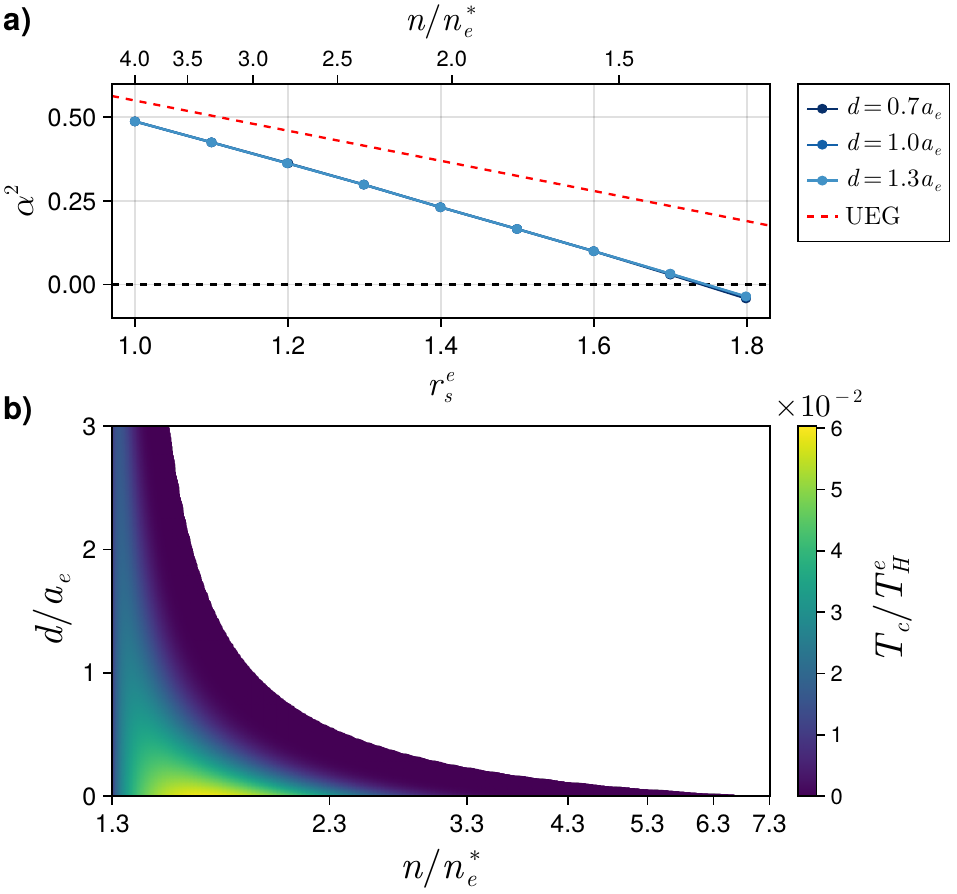}
    \caption{\textbf{a)} The Hartree-Fock calculation of $\alpha^2$ in the EL-HC phase at $m_h/m_e = 10$, fixing $m_h = m_0$. Across different interlayer spacings, there is an approximately linear relationship between $\alpha^2$ and $r_s^e$. Also shown is the analytic relation for the Hartree-Fock uniform electron gas. \textbf{b)} The calculated BCS critical temperature for $m_h/m_e = 10$ in units of the electron Hartree temperature. $n^*_e$ is the monolayer electron crystallization density. Intermediate densities and low interlayer distances give the highest $T_c$. For very high densities, superconductivity does not occur at any interlayer spacing.}
    \label{fig:superphases}
\end{figure}

We can estimate $\alpha$ in the EL-HC phase by calculating the compressibility of the bilayer with Hartree-Fock. Across a range of interlayer spacings, we find a linear relationship between $\alpha^2$ and $r_s^e$, as shown in Fig.~\ref{fig:superphases}. This matches known behavior of interacting electron gasses, with negative values of compressibility occurring in both the Fermi liquid and Wigner crystal phases \cite{chenskii_influence_1980, bello_density_1981, tanatar_ground_1989, pines_theory_1994, eisenstein_negative_1992, shapira_thermodynamics_1996, ilani_unexpected_2000,waintal_quantum_2006, shklovskii_half-century_2024}. For larger $r_s$ however, contributions to the capacitance and compressibility beyond mean-field emerge \cite{skinner_capacitance_2010, loth_non-mean-field_2010}. See the Supplemental Material for more details \cite{supp}. 

An interesting consequence of this behavior is the enhancement of the superconducting critical temperature at intermediate densities. On the one hand, Eq.~\ref{eq:super} predicts an enhancement of Cooper pairing at small densities. This primarily results from the reduction of the effective interlayer distance $\beta$, leading to stronger electron-plasmon coupling. 
Also, $\alpha$ vanishes as one approaches the crystal phase, which enhances the pairing interaction. However, the vanishing of $\alpha$ decreases the Debye temperature, overall lowering $T_c$. This competition suggests that the highest critical temperature will be found at intermediate densities. In fact, the EL-HC phase is also lost at sufficiently small densities---either due to Wigner crystallization or exciton condensation in both layer---putting a lower limit on the density. Similarly, while the maximum critical temperature always occurs at $d=0$, the system may become an exciton condensate. A plot of $T_c$ for the $m_h/m_e =10$ case is given in Fig.~\ref{fig:superphases}. 

Concretely, Hartree-Fock estimates $\alpha \approx 0.3$ near $r_{s}^e = 1.6$ and $m_h/m_e = 10$. If $m_h= m_0$ (the bare mass) and $\epsilon_r =5$, this corresponds to a carrier density $n\approx 1.8 \cdot 10^{12} \text{ cm}^{-2}$ and electron Hartree temperature $T_H^e = 1.26 \cdot 10^3$ K. The Debye temperature is $T_D \approx 94$ K, owing to the small electron mass, with $T_c \simeq 10$~K at $d = a_e$. For comparison, if we take $m_h/m_e = 2000$ but set $m_e = m_0$ and $\epsilon_r = 1$---mirroring hydrogen---this results in $T_c \simeq 180$ K. Increasing $m_h$ affects $T_c$ due to both the change of $T_D$ and the dimensionless pairing interaction strength $g_0 V$; the latter depends on 
the mass ratio $m_h/m_e$ through the effective interlayer distance $\beta$ and the effective acoustic plasmon velocity $\alpha$. 



Our model predicts superconductivity across a range of interlayer distances and densities, with $T_c$ highest at intermediate densities. These conclusions follow from physical considerations of the behavior of the dimensionless plasmon velocity and other dimensionless parameters describing our system. Thus, while we have used multiple approximations in our calculation of specific values of $T_c$, we expect our general predictions to be robust. Note that our current accounting of Umklapp processes is an underestimate, suggesting that superconductivity could be even more generic.

\textit{Conclusions}---In this Letter we have explored how equal-density mass-asymmetric electron-hole bilayers are a rich platform for realizing exciton superfluids, Wigner crystals, and artificial metallic hydrogen. 
At large mass anisotropy, we find at least two quantum phase transitions upon reducing the density, driven by the crystallization of holes and the formation of an exciton condensate respectively. 

Focusing on the intermediate EL-HC phase, we have further shown that acoustic plasmons mediates an effective pairing interaction across a range of densities and interlayer distances, thus leading to superconductivity from Coulomb interactions in a density-balanced electron-hole bilayer. 
This prediction of plasmon-mediated superconductivity, which is strongest at intermediate densities and mass asymmetry, contrasts the exciton superfluid predicted at low densities and small mass asymmetry \cite{zhu_2d_2025}. 

More accurate predictions of both the phase boundaries and superconducting critical temperature can be obtained from advanced theoretical and numerical methods. Eliashberg theory would give more accurate estimates for $T_c$ in the strong-coupling regime (small $\alpha$). Meanwhile, variational Monte Carlo with expressive neural-network wavefunctions can be used to determine the full phase diagram in a unified framework and quantitatively predict phase transitions \cite{cassella_discovering_2023-1, geier_self-attention_2025-1, luo_pairing-based_2023}. A more accurate treatment of liquid and crystal phases may also refine our calculation of $T_c$, as discussed in the Supplemental Material.

As the pairing interaction in our theory of superconductivity is mediated by acoustic plasmons, it is worth exploring plasmon-mediated superconductivity in related settings. This includes the possibility of superconductivity when the light electrons and heavy holes are both in Fermi liquid regimes. Also of interest is superconductivity in bilayer or multiband electron systems, where conducting electrons interact with a lattice of localized electrons and the coupled charge fluctuations mediate (possibly non-$s$-wave) superconductivity from repulsive interactions  \cite{slagle_charge_2020, crepel_spin-triplet_2022}.  

Experimental verification of these predictions could be accomplished in 
TMD-bilayer graphene heterostructures---exploiting the heavy mass of holes in TMDs and the light mass of Dirac fermions in graphene. 
We also note that an external moir\'e potential in the hole layer may help stabilize the hole crystal and facilitate experimental observation of the EL-HC phase and superconductivity. We hope this work motivates experimental study of mass-asymmetric electron-hole bilayers.

\begin{acknowledgments}

\textit{Code Availability}---The code used to perform the Hartree-Fock and Hartree-Fock-Bogoliubov calculations for this paper is available at \cite{code}.

\textit{Acknowledgments}---We thank Aidan Reddy for useful discussions and Boris Shklovskii for insightful comments. This work was supported by a Simons Investigator Award from the Simons Foundation. L.M.N was supported by a Columbia University I.I. Rabi Scholarship.  
\end{acknowledgments}

\bibliography{permanent_references}

\widetext
\setcounter{equation}{0}
\setcounter{figure}{0}
\setcounter{table}{0}
\setcounter{page}{1}
\makeatletter
\renewcommand{\theequation}{S\arabic{equation}}
\renewcommand{\thefigure}{S\arabic{figure}}
\begin{center}
\textbf{\large Supplemental Material: Prediction of superconductivity in mass-asymmetric electron-hole bilayers}
\end{center}
\section{Hartree-Fock Frameworks}
Our model Hamiltonian is 
\begin{equation}\label{eq:hamil}
    \begin{split}
        H &= \sum_{{\bf k}, s} \frac{\hbar^2 {\bf k}^2}{2m_e} a^\dagger_{{\bf k}, s}a_{{\bf k}, s} + \sum_{{\bf k}} \frac{\hbar^2 {\bf k}^2}{2m_h} b^\dagger_{{\bf k}}b_{{\bf k}} + \frac{1}{2}\sum_{\substack{{\bf k}, {\bf k}', {\bf q}\\s, s'}} V_{0}({\bf q})a^\dagger_{{\bf k}+{\bf q}, s} a^\dagger_{{\bf k}'-{\bf q}, s'} a_{{\bf k}', s'} a_{{\bf k}, s}\\
        &\phantom{==} + \frac{1}{2} \sum_{\substack{{\bf k}, {\bf k}', {\bf q}}} V_{0}({\bf q})b^\dagger_{{\bf k}+{\bf q}} b^\dagger_{{\bf k}'-{\bf q}} b_{{\bf k}'} b_{{\bf k}}- \sum_{\substack{{\bf k}, {\bf k}', {\bf q}\\s}} V_{1}({\bf q})a^\dagger_{{\bf k}+{\bf q}, s} b^\dagger_{{\bf k}'-{\bf q}} b_{{\bf k}'} a_{{\bf k}, s}
    \end{split}
\end{equation}
where $a_{{\bf k}, s}$ is the annihilation operator for an electron of spin $s$, $b_{{\bf k}}$ is the annihilation operator for a spin-polarized hole, $V_0({\bf q}) = 2\pi e^2 /(\epsilon_r q A)$ is the 2D intralayer Coulomb interaction, and $V_1({\bf q}) = 2\pi e^2 e^{-qd}/(\epsilon_r  qA)$ is the interlayer interaction. As is standard, we set the $q=0$ component of these interactions to 0 (as discussed below, this is not strictly necessary, but is the most relevant scheme for our needs). The code for these calculations is available at \cite{code}. 

Using the method of \cite{dai_strong-coupling_2024}, we can perform direct-minimization Hartree-Fock to study this Hamiltonian. Specifically, we directly minimize the Hartree-Fock energy functional
\[E[D] = \sum_{\alpha, \bf k} \frac{\hbar^2 {\bf k}^2}{2m_\alpha} D^\alpha_{\bf k,\, \bf k} + \frac{1}{2} \sum_{\substack{\bf k, \bf k', \bf q\\\alpha, \beta}} \left[V^{\alpha \beta}({\bf q}) D^{\alpha}_{\bf k + \bf q, \bf k} D^{\beta}_{\bf k' - \bf q, \bf k'}- \delta^{\alpha \beta}V_0({\bf q}) D^{\alpha}_{\bf k + \bf q, \bf k'} D^{\beta}_{\bf k' - \bf q, \bf k}  \right],\]
where $\alpha = \{e_\uparrow, e_\downarrow, h\}$ is a species index and $V^{\alpha\beta}$ is the appropriate intra/interlayer Coulomb interaction. $D^\alpha_{\bf k\, \bf k'} = \langle c^{\alpha\, \dagger}_{\bf k} c^\alpha_{\bf k'}\rangle$ is the density matrix subject to appropriate constraints on positivity and particle number $N_\alpha$. These are automatically implemented by expressing $[D^\alpha]^T = C^\alpha (C^{\alpha\,\dagger} C^\alpha)^{-1}  C^{\alpha\,\dagger}$ for $C^\alpha_{{\bf k}, n}$ an $N_\alpha$-column matrix collecting the single-particle states. As we expect the formation of a triangular Wigner crystal phase, we model this Hamiltonian on a periodic triangular lattice (i.e. the $\bf k$ are taken from the reciprocal triangular lattice). A cutoff for the $\bf k$ points is chosen so that the occupations $D^\alpha_{\bf k,\bf k}$ are less than $1\%$ for $\bf k$ at the cutoff. 

To best capture the features of different phases, we perform both unrestricted (with no constraints) and restricted (where the $e_{\uparrow/\downarrow}$ density matrices are forced equal) Hartree-Fock calculations. Specifically, we use unrestricted Hartree-Fock in the Wigner crystal phase, as the electron spin-orbitals are necessarily distinct in this regime. In the electron liquid regime, we perform both unrestricted and restricted Hartree-Fock to mitigate the formation of charge density waves \cite{overhauser_spin_1962} and confirm they produce similar results. 

While our results show the formation of the electron Wigner crystals around $r_s \simeq 2$---in rough agreement with the Hartree-Fock literature \cite{trail_unrestricted_2003}---more sophisticated methods and experimental work indicate that the true critical crystallization threshold is closer to $r_s^e \simeq 31$ due to the strong electron correlations of the liquid state \cite{waintal_quantum_2006, ludwig_crystallization_2007, schleede_phase_2012, li_imaging_2021}. To bridge this gap, we normalize our findings to a crystallization threshold $r_s^*$, which may either correspond to the Hartree-Fock or physical value. 

We also consider a BCS-type pairing ansatz---which is equivalent to a homogeneous Hartree-Fock-Bogoliubov theory \cite{luo_pairing-based_2023}---to better explore the condensate regime. To do so, we assume the electron spin is polarized so that $a_{{\bf k}, s}$ may be reduced to $a_{{\bf k}}$, and consider the ansatz
\begin{equation}
    |{\Omega}\rangle = \prod_{{\bf k}} (\cos\theta_{{\bf k}} + \sin\theta_{{\bf k}} a^\dagger_{{\bf k}}b^\dagger_{-{\bf k}}) |{0}\rangle.
\end{equation}
As the particle number is not fixed in this ansatz, we instead minimize the free-energy $F = H-\mu N$, where $2\mu$ can be viewed as the exciton chemical potential and can be tuned to reproduce the particle densities used in multi-species Hartree-Fock. Explicitly, we have
\begin{equation}
    F = \sum_{\bf k} \left (\frac{\hbar^2 {\bf k}^2}{2\tilde m} - 2\mu_E\right )\sin^2(\theta_{\bf k})- \sum_{{\bf k}, {\bf k}'} V_0({\bf k} - {\bf k}') \sin^2(\theta_{\bf k}) \sin^2(\theta_{{\bf k}'}) - \sum_{{\bf k}, {\bf k}'} V_1({\bf k} - {\bf k}') \frac{\sin(2\theta_{\bf k}) \sin(2\theta_{{\bf k}'})}{4}
\end{equation}
where $\tilde m = m_e m_h/(m_e+m_h)$ is the reduced mass of the electron-hole pair. This expression is almost a quadratic form in $n_{\bf k} = \sin^2(\theta_{\bf k})$, but the interlayer coupling term introduces a nonlinearity. Any number of methods can be used to optimize this free energy and thus deduce the occupations. The momentum cutoff is chosen so that $n_{\bf k}<10^{-3}$ for all $\bf k$ at the cutoff. 

\section{Details for Pairing Potential}
We first state some standard relations:
\[k_F = \frac{\sqrt{2}}{r_s^e a_e} = \sqrt{2\pi n}\qquad\text{and}\qquad k_D = \sqrt{2} k_F.\]
Next, the Bohm-Staver approximation to the sound velocity is obtained by assuming the compressibility of a material is that of a free electron gas, while the mass density comes solely from the ions. Recalling the 2D Fermi gas energy relation $E_F = N\epsilon_F/2$, we thus get
\[B_0 = A \frac{\partial^2 E_F}{\partial A^2} = n \epsilon_F \qquad \text{and}\qquad v_0^2 = \frac{B_0}{\rho} = \frac{\epsilon_F}{m_h},\]
where $B$ is the thermodynamic bulk modulus, and is related to the electronic compressibility by $\kappa = \partial n /\partial \mu = n^2/B$.

\begin{figure}[t]
    \centering
    \includegraphics[width=0.6\linewidth]{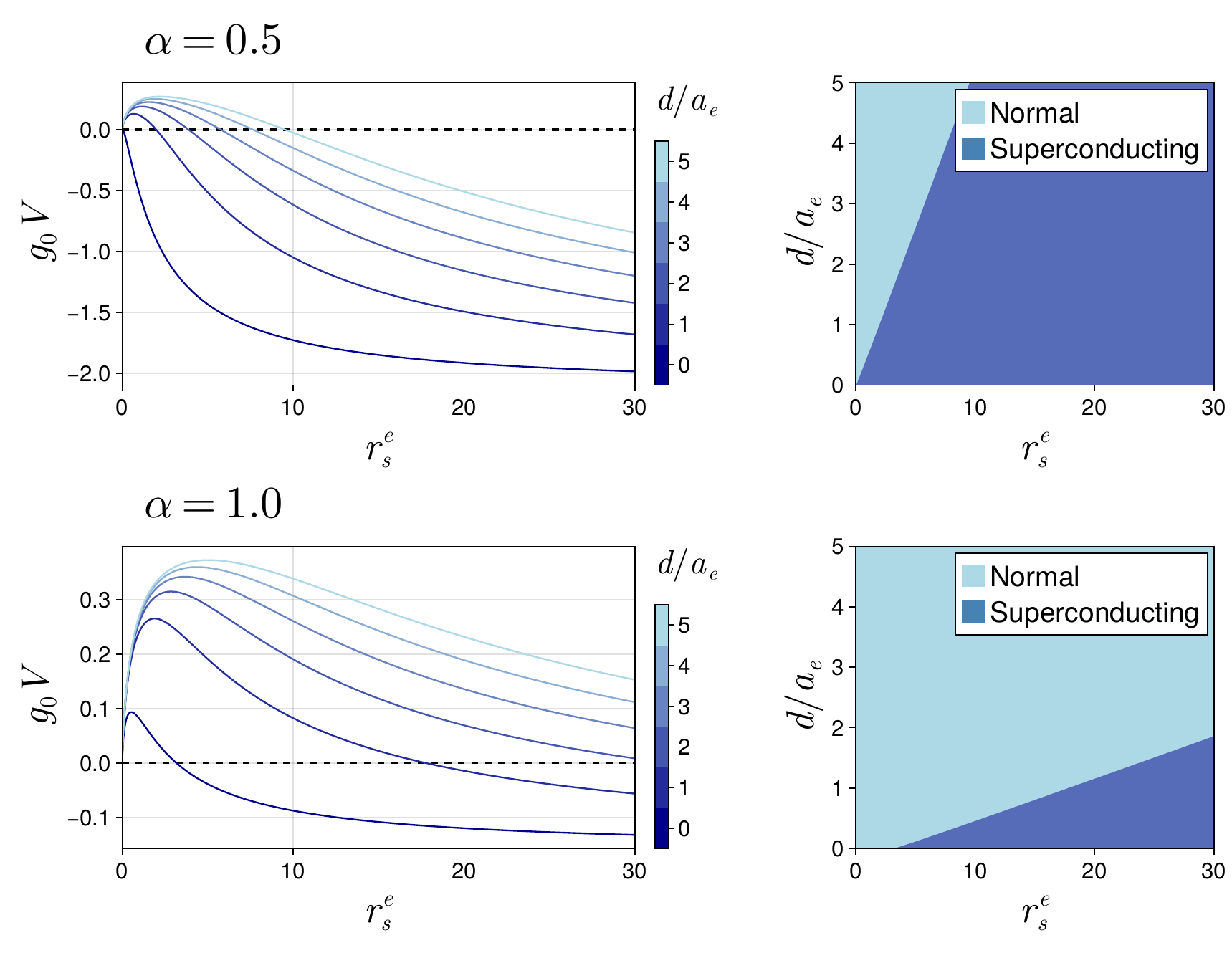}
    \caption{Calculation of the average effective Cooper pair potential and the onset of superconductivity for fixed values of $\alpha$. Because of the finite interlayer spacing, electron-plasmon interactions are strongly suppressed for high densities.}
\end{figure}

Moving on to the pairing potential, we consider two parts contributing to the net electron-electron interaction. The direct repulsion takes the standard form of Eq.~\ref{eq:V_base} for a 2D screened Coulomb interaction. For the plasmon-mediated contribution, we start with the electron-plasmon coupling which, for a general Coulomb coupling $U_1$ between electron and holes, looks like \cite{bruus_many-body_2004}
\[M_{\bf q}({\bf k} - {\bf k}') = i \sqrt{\frac{n\hbar}{2 m_h \omega_{\bf q}}}\; |{\bf k} - {\bf k}'|\; U_1({\bf k} -{\bf k}').\]
In the case of the bilayer system, we take $U_1$ to approximately be
\[U_1({\bf k}) = \frac{2\pi e^2}{\epsilon_r}\frac{e^{-d|{\bf k}|}}{|{\bf k}| + k_s},\]
which roughly captures the effects of both screening and interlayer separation, assuming intralayer screening dominates. Eq.~$\ref{eq:V_avg}$ then follows by applying second-order perturbation theory to eliminate the intermediate plasmon interaction and determine the full effective electron-electron interaction. Lastly, to get from \ref{eq:V_avg} to \ref{eq:super} is standard algebra, where $\theta$ ultimately indicates half the angle between $\bf k$ and ${\bf k}'$ (i.e. $|{\bf k} - {\bf k}'| = 2k_F \sin(\theta)$). Of note is that the transition from the normal to Umklapp regime occurs at $\theta = \pi/4$. 

\section{Capacitance Energy and Compressibility}
To calculate $\alpha$, we numerically calculate the bulk modulus of the system with restricted Hartree-Fock, as the system is in the EL-HC phase over the parameter regime of interest. By evaluating the total energy $E$ at areas $A-h, A, A+h$, we can numerically calculate the bulk modulus as $B = A (E_{A+h} + E_{A-h} - 2E_A)/h^2$. Very high convergence criteria must be set for the numerical stability of this calculation. We confirmed our results were consistent using different values of $h$ and slightly shifted values of $A$. 

\begin{figure}
    \centering
    \includegraphics[width=0.6\linewidth]{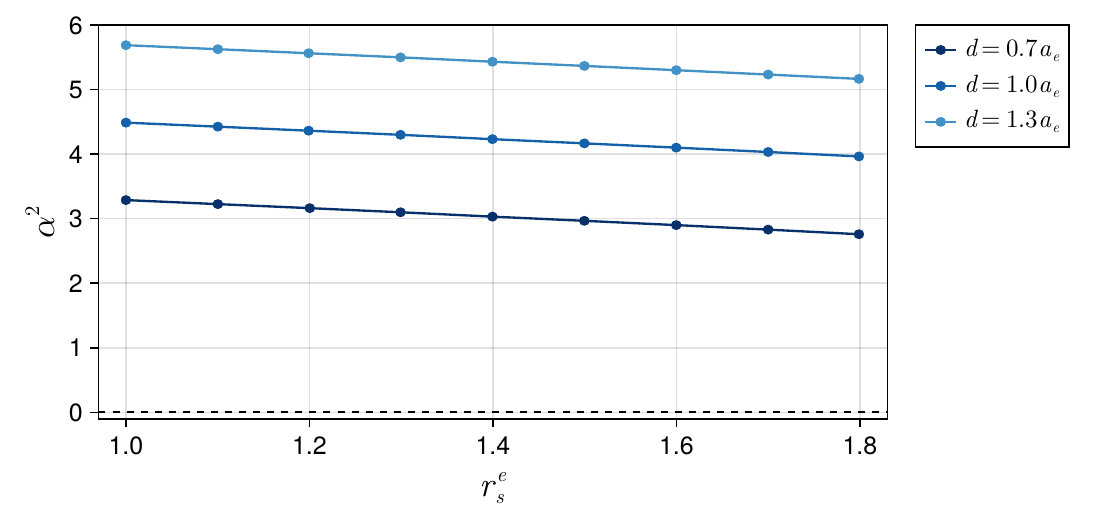}
    \caption{Including the capacitance energy effectively modifies $\alpha^2 \to \alpha^2 + 4 d/a_e$ compared to Fig.~\ref{fig:superphases}. As a result, the compressibility of the system is always positive for the $r_s$ and $d$ considered.}
    \label{fig:compcap}
\end{figure}

Because our system is charge-neutral, the choice to set the $V({\bf q}=0)=0$ in the Coulomb interaction is not strictly necessary. Doing so effectively ignores a geometric capacitance energy $E_{C} = 2\pi e^2 N^2 d/(\epsilon_r A) = 4E_F \,d/a_e $. When comparing different Hartree-Fock ground state energies under the same circumstances, it is unnecessary to include this geometric capacitance energy, as it is a constant shift independent of the electronic phase. However, since capacitance energy does enter in the expression for thermodynamic compressibility, including it would suggest values for $\alpha = \sqrt{B/B_0}$ far exceeding 1 as shown in Fig.~\ref{fig:compcap}.

For the purpose of calculating $\alpha$ however, it is correct to ignore this geometric capacitance energy. To see this, recall that $\alpha$ is defined as the ratio between the plasmon velocity and the natural Bohm-Staver velocity. To estimate the velocity of the acoustic plasmon mode in the electron–hole bilayer, we employ the standard relation $v_s = \sqrt{B/\rho}$. Note however that this relation is given by the long-wavelength elasticity theory of a macroscopically homogeneous charge-neutral medium: the linearized equations of motion for local density fluctuations are a wave equation with sound velocity $v_s$. Crucially, this derivation assumes that the energy cost of density modulations arises from local interactions, and that these local modulations do not cause any net charge fluctuations.

In the bilayer system, this distinction is essential. The geometric capacitance energy represents the global, macroscopic energy cost associated with charging two separated layers. While this homogeneous energy contributes to the bare compressibility when varying area, it does not enter the dynamics of local density fluctuations that leave the global charge density unchanged. The bulk modulus entering into the sound velocity should therefore be understood as arising from the \textit{inhomogeneous} or \textit{local} energy $B=A\; \partial^2\! E_{\text{inhom}}/\partial^2\! A$ under conditions of overall charge neutrality.

Returning to the microscopic picture, this is also clear. Both standard phonons and interlayer acoustic plasmons correspond to collective excitations with finite wavevector $\mathbf{q} \neq 0$, representing local density fluctuations with no net charge fluctuation. As a result, they do not couple to the geometric capacitance energy $E_C$, which is sensitive only to the $\mathbf{q}=0$ component of the charge density. Instead, these modes couple to \textit{local} Coulomb interactions, i.e., to the finite-$\mathbf{q}$ components of the interaction potential. The restoring force governing their dynamics is therefore controlled by the compressibility induced by finite $\bf q$ interactions, but not the macroscopic charging energy. 

To estimate the plasmon velocity in the regime of small $d$, it is thus correct to use the standard sound velocity relation with the geometric capacitance excluded from the bulk modulus. We remark that the capacitance energy becomes non-geometric for larger $r_s$ beyond the range of Hartree-Fock's applicability \cite{skinner_capacitance_2010, loth_non-mean-field_2010}, however we do not expect superconductivity in this regime.


Next, we note that Hartree-Fock gives a prediction of $\alpha^2$ for the interacting electron gas. If one treats the Fock interaction as a perturbation to the free Fermi gas, then the first-order perturbation to the energy is \cite{giuliani_quantum_2005}
\[E_{HF} = \frac{1}{2r_s^2} - \frac{4\sqrt{2}}{3\pi r_s}\]
in Hartree units. Starting with this expression for the energy, $\alpha^2$ has the exact value
\[\alpha^2 = \frac{B_{HF}}{B_{0}} = 1 - \frac{\sqrt{2}}{\pi} r_s. \]

\begin{figure}
    \centering
    \subfloat{\includegraphics[width=0.49\linewidth]{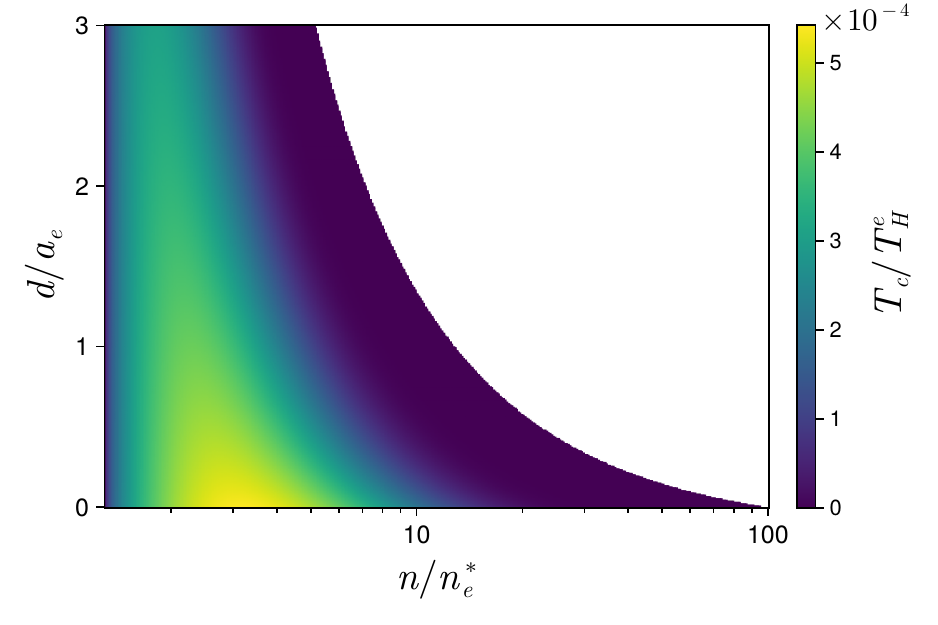}}
    \hfill
    \subfloat{\includegraphics[width=0.49\linewidth]{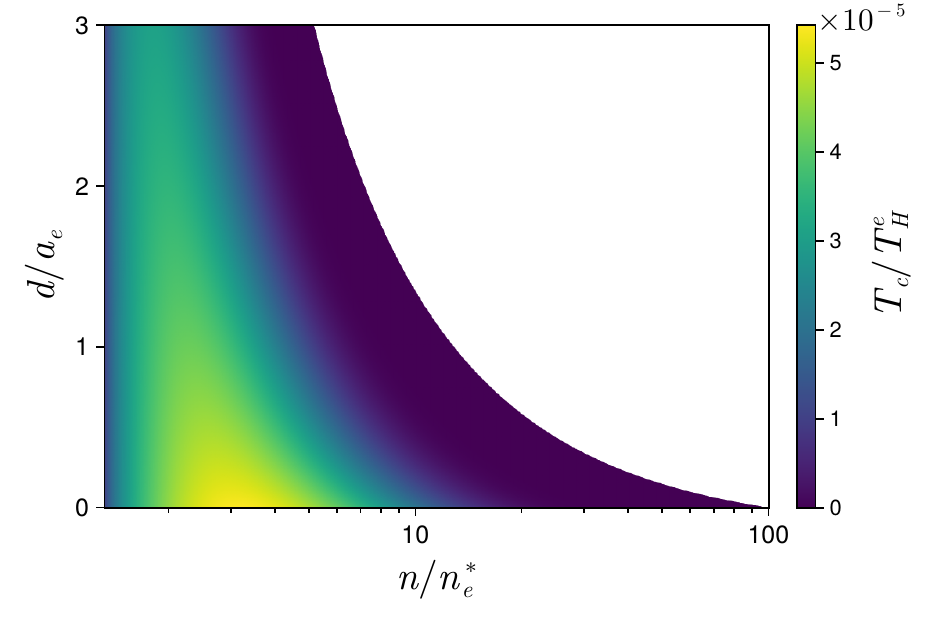}}
    \caption{\textbf{Left:} The calculated BCS critical temperature for $m_h/m_e = 10$---the same scenario as in the main text---but with $r_s^* = 31$ as predicted by other numerical methods. The qualitative features are maintained, though superconductivity occurs in a much broader region and at much lower temperatures.  \textbf{Right: } The calculated BCS critical temperature for $m_h/m_e = 1000$. If $m_e = m_0$ and $\epsilon_r = 1$ (in analogy to solid hydrogen), this gives $T_c \simeq 16$ K. }
    \label{fig:Tc_rescaled}
\end{figure}

Lastly, Hartree-Fock is unable to capture the physical liquid-crystal transition around $r_s^{*} \simeq 31$, predicting much lower crystallization densities $r_s^{*} \simeq 2$. As a result, our calculations for the compressibility of the EL-HC phase are restricted to this nonphysical parameter regime. Remarkably, the Hartree-Fock compressibility of the pure electron gas is still in some agreement with experiment beyond this range \cite{ilani_unexpected_2000, tanatar_ground_1989, shklovskii_half-century_2024}. For sufficiently high mass ratios and sufficiently small $d$ though, the negative compressibility we calculate at moderate $r_s^e$ may not occur, as physically the system approaches a stable ``atomic metal''. To alleviate this issue, we consider a physically motivated rescaling of $\alpha^2$ so that it is a function of the ratio $r_s^e/r_s^{*}$ i.e. the ratio between the electron Wigner-Seitz radius to the critical value for monolayer crystallization. Our Hartree-Fock calculation then gives $\alpha^2$ in the case $r_s^* = 2$. By using a more accurate value of $r_s^* = 31$ but otherwise keeping the same trend, we can partially address the limitations of our pure Hartree-Fock calculation. Corresponding values of $T_c$ are shown in Fig.~\ref{fig:Tc_rescaled}.

\section{Hartree-Fock Phase Criteria}
\begin{figure}[t]
    \centering
    \includegraphics[width=0.95\linewidth]{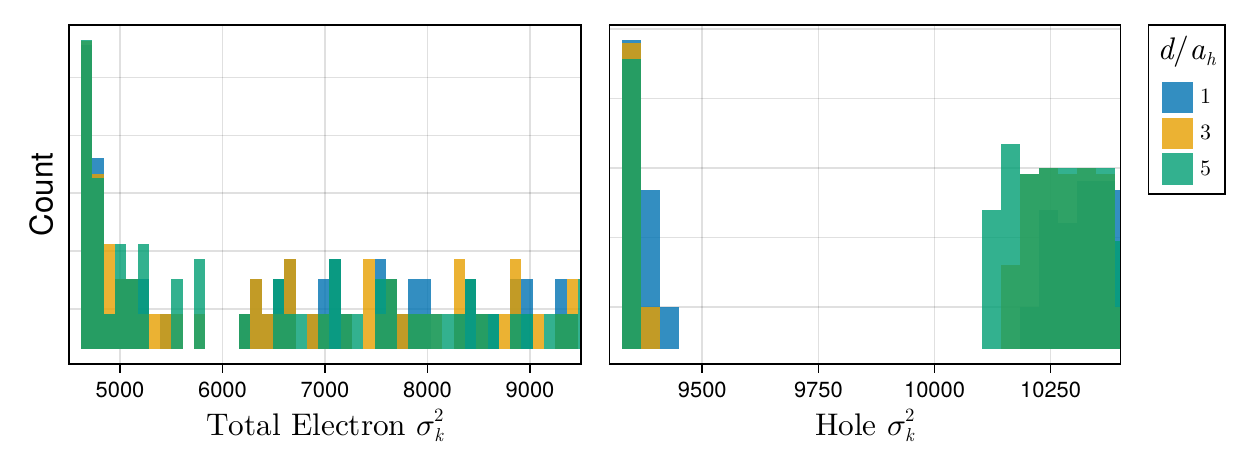}
    \caption{A histogram showing $\sigma_k^2$ for all points in the phase diagrams for Fig.~\ref{fig:phases}. Across the range of interlayer distances, densities, and mass ratios used to create the phase diagrams, choosing cutoffs of $\sigma_{\bf k}^2 \sim 6000$ and $\sigma_{\bf k}^2 \sim 10000$ for electrons and holes respectively separates the crystal and liquid phases. The precise values are arbitrary---depending on parameters like particle number---but provide clean separation for our use.}
    \label{fig:criteria}
\end{figure}

To distinguish the Wigner crystal and Fermi liquid phases, we rely on the fact that the liquid state has a well-defined Fermi surface and, thus, a more narrow distribution in momentum space. In particular, by calculating the scale-independent momentum-space variance
\[\sigma_{k}^2 = \sum_{\bf k, s} (L{\bf k})^2 \langle a^\dagger_{\bf k, s} a_{\bf k, s}\rangle,\]
we obtain an approximate heuristic distinguishing the two phases. Though not as exact as previous studies using different variational wavefunctions for the two phases, we find numerically that $\sigma_{k}^2$ does provide a clean separation between the two phases, especially in the spin-polarized layer, as shown in Fig.~\ref{fig:criteria}.

\newpage
\section{Additional Hartree-Fock Results}
Below we display a variety of real-space and momentum-space densities produced by multi-species Hartree-Fock. Unless otherwise specified, color bars show densities with the unit cell area normalized to unit. The full charge density plots show 4 unit cells, while the number density plots show a single unit cell. All calculations were performed with $m_h = m_0$ and with 36 holes.
\begin{figure}[ht]
    \centering
    \includegraphics[width=0.95\linewidth]{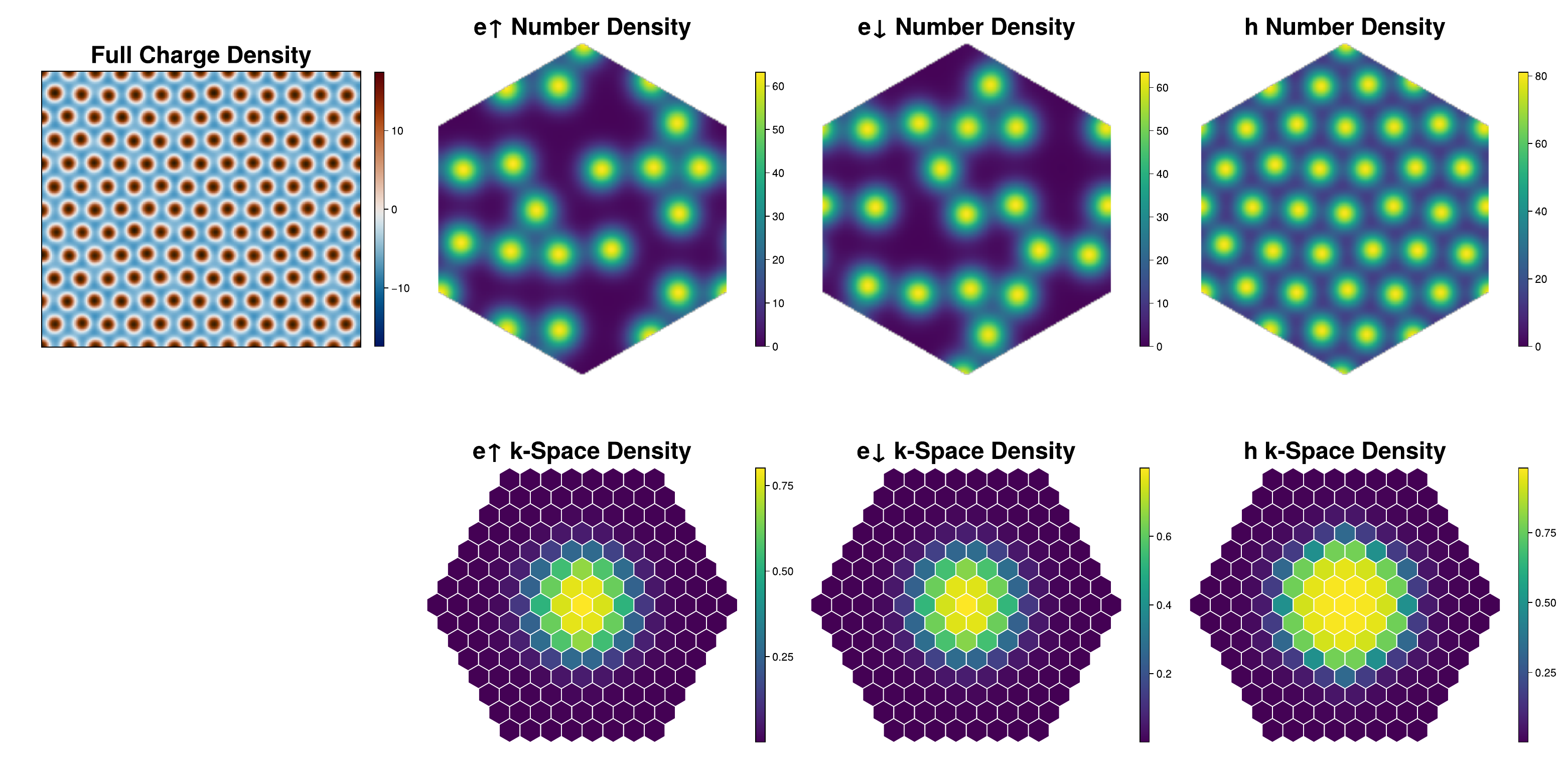}
    \caption{The coupled crystal state with a triangular lattice initialization for $d=1\, a_h$ and $m_h/m_e = 8$ at $n/n^* = 0.1$. The exact triangular symmetry is broken by the lack of symmetry in the spin texture.}
\end{figure}

\begin{figure}[ht]
    \centering
    \includegraphics[width=0.95\linewidth]{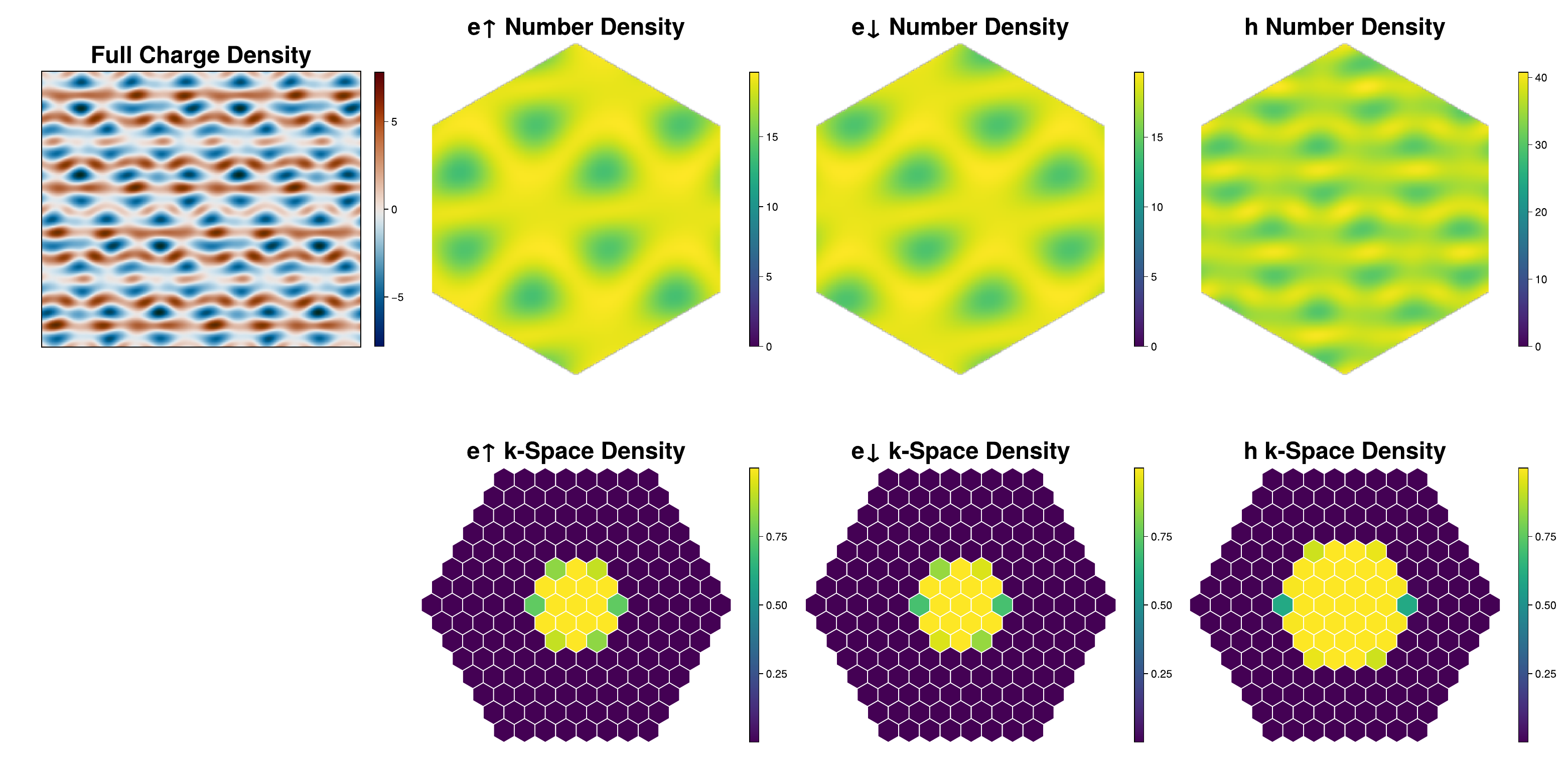}
    \phantom{ }\vspace{5ex}
    \includegraphics[width=0.95\linewidth]{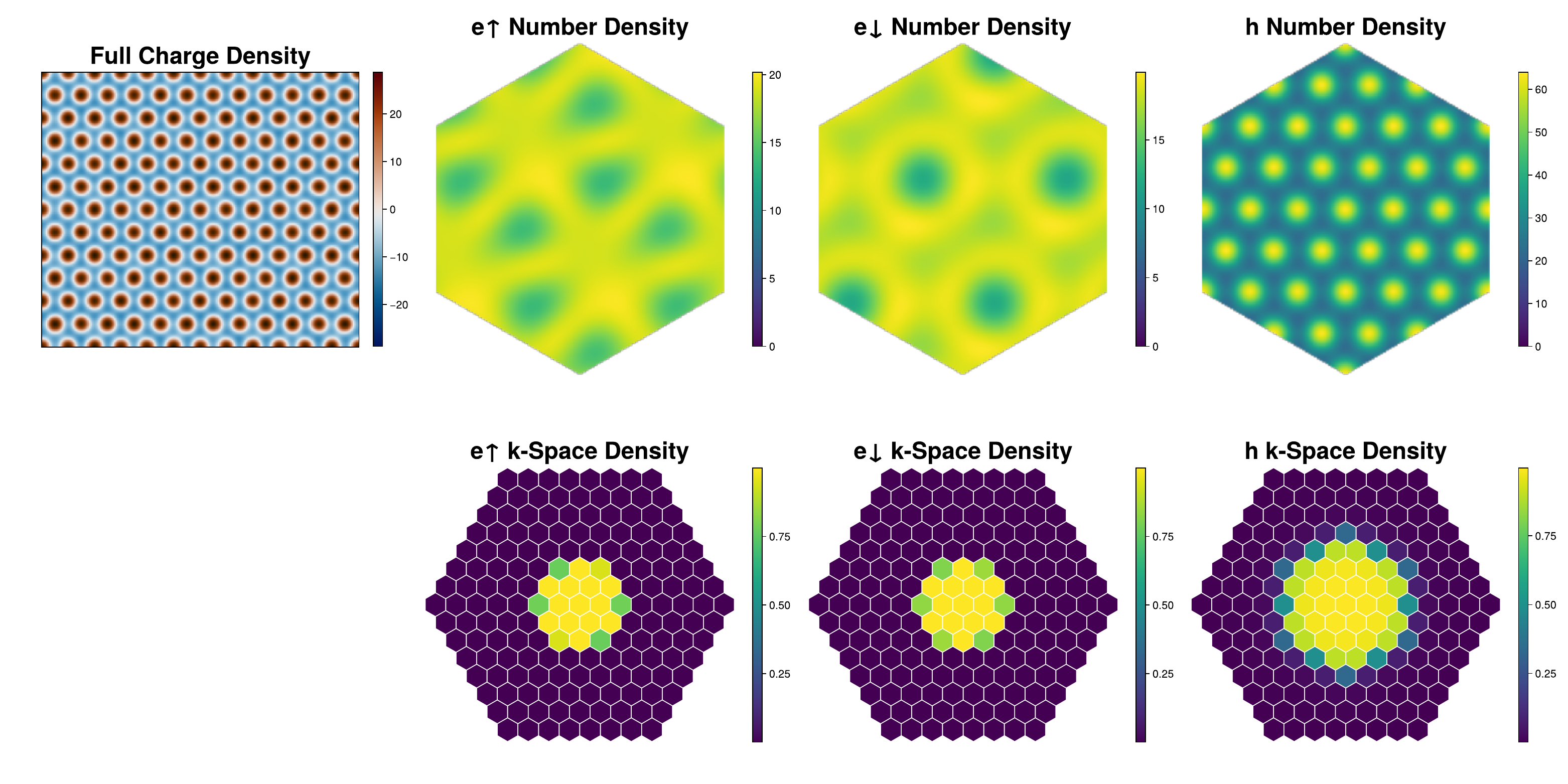}
    \caption{The fully liquid and EL-HC states in unrestricted Hartree Fock with a triangular lattice initialization for $d=3\, a_h$ and $m_h/m_e = 10$. The exact triangular symmetry is broken by the lack of symmetry in the spin texture. The fully liquid phase (upper) corresponds to $n/n^* = 1.3$, while the EL-HC phase (lower) corresponds to $n/n^*=0.8$. }
\end{figure}

\begin{figure}
    \centering
    \includegraphics[width=0.95\linewidth]{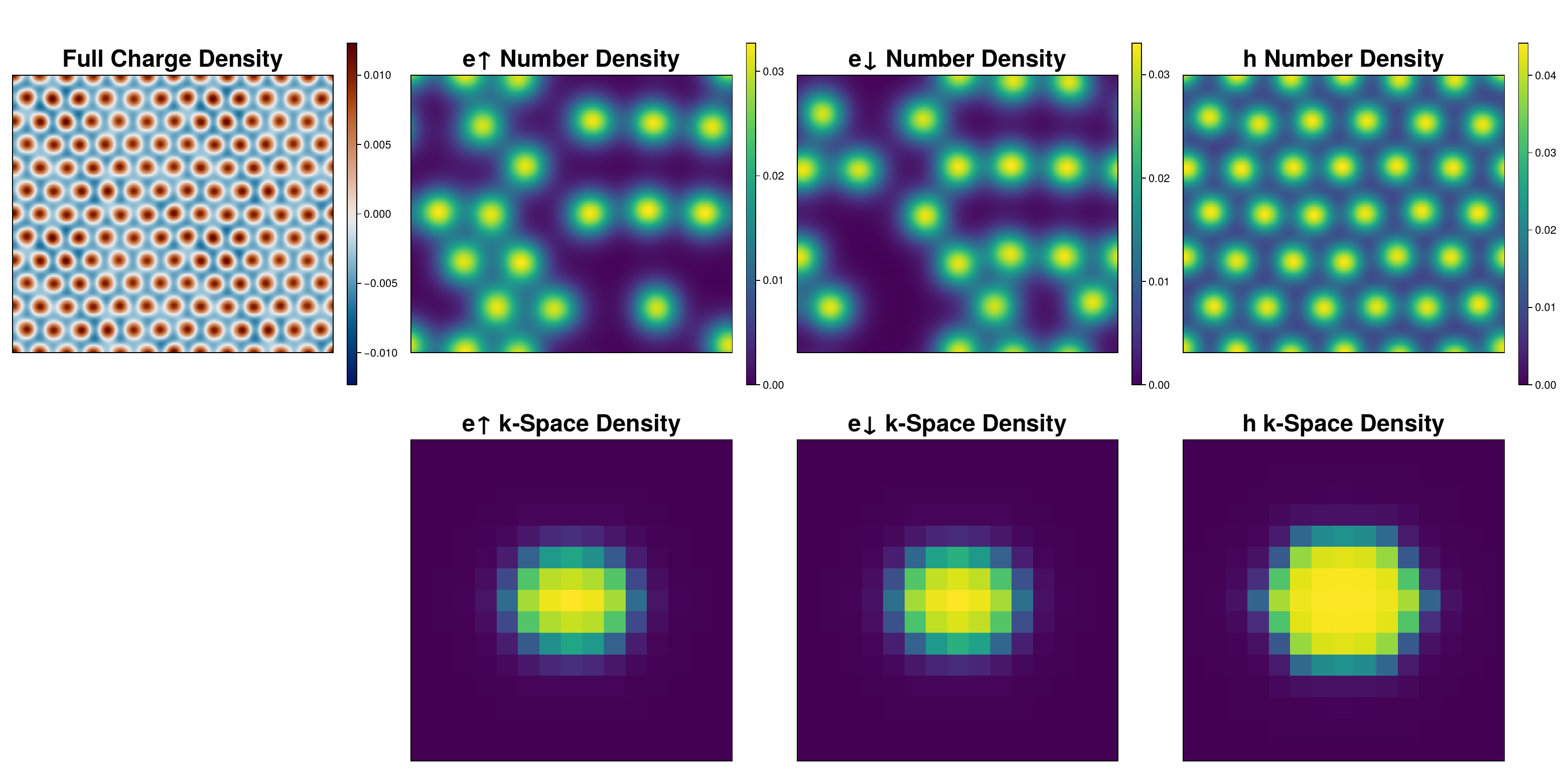}
    \caption{A triangular WC is formed even when not explicitly imposing triangular symmetry, but rather using a large rectangular box. Here, $d=1\;a_h$, $m_h/m_e = 2$, and $n/n^* = 0.8$. The color bars show densities with the unit cell area left in $1/a_h^2$ units.}
\end{figure}

\begin{figure}
    \centering
    \includegraphics[width=0.95\linewidth]{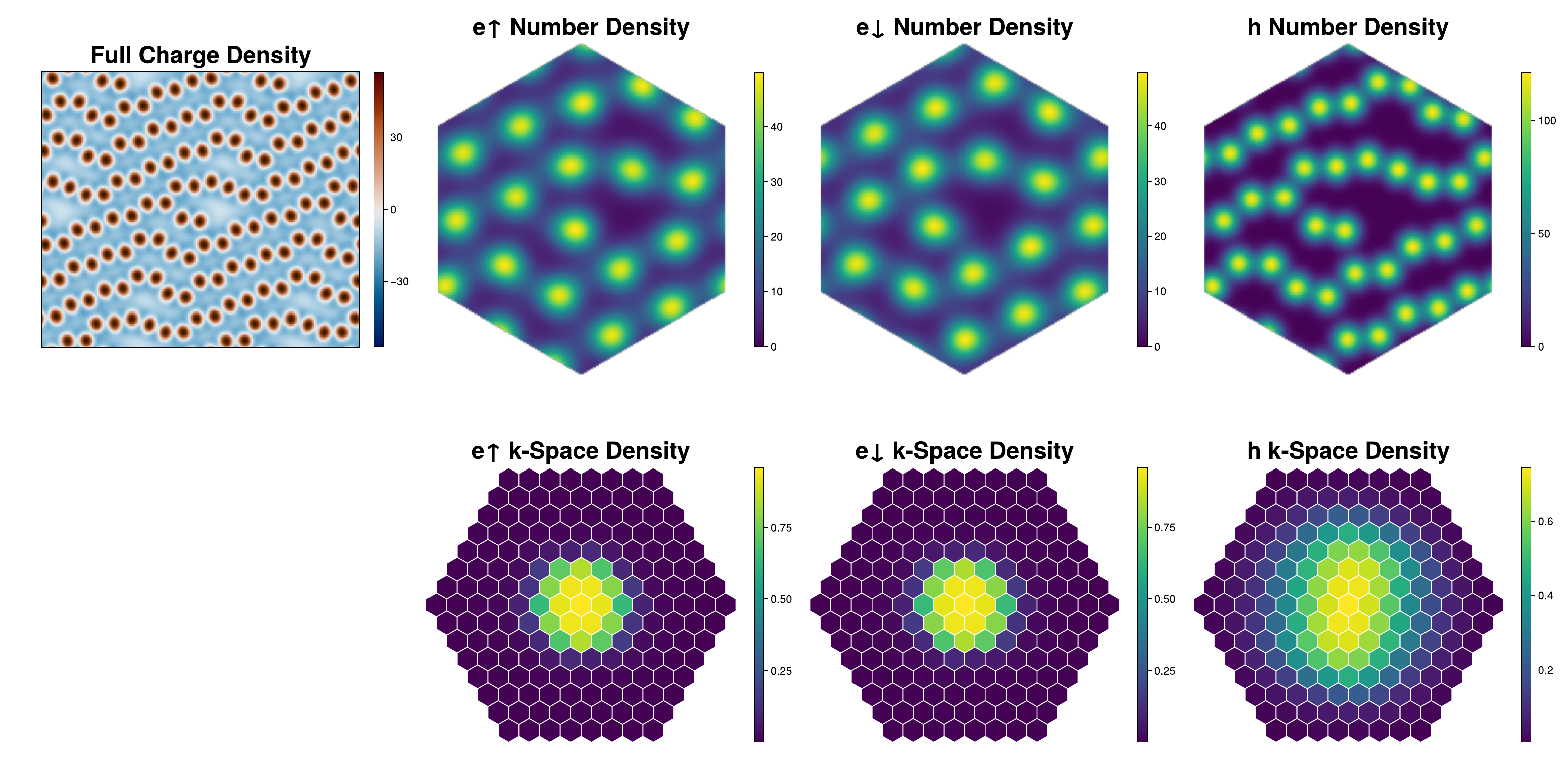}
    \caption{Interestingly, the Wigner crystal can already begin melting in standard Hartree-Fock. Pictured above are HF results from $d=1\;a_h$, $m_h/m_e = 8$, and $n/n^* = 0.1$. When the density is lowered sufficiently, the dipole interactions between interlayer excitons are not sufficient to maintain a clean lattice structure, and bi-exciton stripes emerge.}
\end{figure}

\end{document}